\documentclass[sigconf, authorversion]{acmart}

\usepackage[linesnumbered,ruled,vlined]{algorithm2e}

\AtBeginDocument{%
  \providecommand\BibTeX{{%
    \normalfont B\kern-0.5em{\scshape i\kern-0.25em b}\kern-0.8em\TeX}}}

\copyrightyear{2020}
\acmYear{2020}
\setcopyright{acmcopyright}
\acmDOI{10.1145/3351095.3372863}

\acmConference[FAT* '20]{Conference on Fairness, Accountability, and Transparency}{January 27--30, 2020}{Barcelona, Spain}
\acmBooktitle{Conference on Fairness, Accountability, and Transparency (FAT* '20), January 27--30, 2020, Barcelona, Spain}
\acmPrice{15.00}
\acmISBN{978-1-4503-6936-7/20/02}

\begin{document}

\title[Predictive Equity Case Study]{Case Study: Predictive Fairness to Reduce Misdemeanor Recidivism Through Social Service Interventions}

\author{Kit T. Rodolfa}
\email{krodolfa@cmu.edu}
\affiliation{%
  \department{Machine Learning Department}
  \institution{Carnegie Mellon University}
}

\author{Erika Salomon}
\email{ecsalomon@gmail.com}
\affiliation{%
  \department{Ctr for Data Science \& Public Policy}
  \institution{University of Chicago}
}

\author{Lauren Haynes}
\email{lnhaynes@gmail.com}
\affiliation{%
  \department{Ctr for Data Science \& Public Policy}
  \institution{University of Chicago}
}

\author{Iv\'an Higuera Mendieta}
\email{ivanhigueram@uchicago.edu}
\affiliation{%
  \department{Ctr for Data Science \& Public Policy}
  \institution{University of Chicago}
}

\author{Jamie Larson}
\email{jamie.larson@lacity.org}
\affiliation{%
  \department{Director, Recidivism Reduction and Drug Diversion Unit (R2D2)}
  \institution{Los Angeles City Attorney's Office}
}

\author{Rayid Ghani}
\email{rayid@cmu.edu}
\affiliation{%
  \department{Machine Learning Department}
  \department{and Heinz College of Information Systems \& Public Policy}
  \institution{Carnegie Mellon University}
}

\renewcommand{\shortauthors}{Rodolfa, Salomon, et al.}

\begin{abstract}
The criminal justice system is currently ill-equipped to improve outcomes of individuals who cycle in and out of the system with a series of misdemeanor offenses. Often due to constraints of caseload and poor record linkage, prior interactions with an individual may not be considered when an individual comes back into the system, let alone in a proactive manner through the application of diversion programs. The Los Angeles City Attorney's Office recently created a new Recidivism Reduction and Drug Diversion unit (R2D2) tasked with reducing recidivism in this population. Here we describe a collaboration with this new unit as a case study for the incorporation of predictive equity into machine learning based decision making in a resource-constrained setting. The program seeks to improve outcomes by developing individually-tailored social service interventions (i.e., diversions, conditional plea agreements, stayed sentencing, or other favorable case disposition based on appropriate social service linkage rather than traditional sentencing methods) for individuals likely to experience subsequent interactions with the criminal justice system, a time and resource-intensive undertaking that necessitates an ability to focus resources on individuals most likely to be involved in a future case. Seeking to achieve both efficiency (through predictive accuracy) and equity (improving outcomes in traditionally under-served communities and working to mitigate existing disparities in criminal justice outcomes), we discuss the equity outcomes we seek to achieve, describe the corresponding choice of a metric for measuring predictive fairness in this context, and explore a set of options for balancing equity and efficiency when building and selecting machine learning models in an operational public policy setting.
\end{abstract}




\maketitle

\section{Introduction}
Some of the most vulnerable populations in the United States struggle with a complex combination of needs, including homelessness, substance addiction, ongoing mental and physical health conditions, and long-term unemployment. For many, these challenges can lead to interactions with the criminal justice system \cite{Hamilton2010PeopleSystem}. Of the millions of people who are incarcerated in jails and prisons each year, more than half have a current or recent mental health problem and inmates are far more likely to have experienced homelessness or substance dependence. In local jails, where 64\% struggle from mental health issues, 10\% were homeless in the year before their arrest (compared to a national average under 1\% \cite{USHUD2017Annual2}), and 55\% met criteria for substance dependence or abuse \cite{James2006MentalInmates}.

By 2005, there were three times as many individuals with serious mental illness in jails and prisons than in hospitals and the per capita number of psychiatric hospital beds in the US had fallen by an order of magnitude over 50 years, suggesting a failure of the community mental health system to meet the needs of this at risk population \cite{FullerTorrey2010MoreStates}. For some of these individuals, the criminal justice system may be their first or primary interaction with social services, but it is particularly poorly suited to address these additional needs. Lacking needed treatment or other interventions, a significant group of individuals cycles through jails and prisons, with the system as a whole failing to appreciably improve their individual outcomes or public safety \cite{Stone1997TherapeuticPolicy,Kondo2000TherapeuticOffenders.,Kutcher2009ProblemsSystem}. The Criminal Justice/Mental Health Consensus Project found widespread dissatisfaction with the lack of resources available in the criminal justice system to address mental illness \cite{Thompson2003CriminalIllness}, and these failings are borne out in the statistics, with recidivism rates for individuals with mental illness reaching as high as 70\% in some jurisdictions \cite{Ventura1998CaseJail}. Likewise, Demleitner \cite{Demleitner2002CollateralOffenders}, argues that the combination of lacking effective treatment and ``collateral restrictions'' (such as restrictions on welfare benefits and employment opportunities) for drug offenders tends to reinforce the cycle of incarceration for people facing substance abuse issues.

Faced with the high costs of incarceration, large jail populations booked with low-level misdemeanor offenses, and poor outcomes for these individuals with complex needs, some communities are turning to restorative justice and pre-trial diversionary programs as an alternative to incarceration in an effort to break this cycle. The design and implementation of these programs is as variable as the needs of the populations they serve, including (for example) mental health services, community service or restitution, substance abuse treatment, and facilitated meetings between victims and offenders. Use of these programs has expanded rapidly over the last two decades \cite{Steadman2005AssessingDisorders} and recent examinations of opportunities to improve outcomes in the criminal justice system have identified wide support for their continued expansion \cite{Thompson2003CriminalIllness}. Evaluations of diversionary programs have generally shown success in reducing the time spent in jail without posing an increased risk to public safety, as well as increasing utilization of social services by individuals with mental health and substance abuse issues \cite{Steadman2005AssessingDisorders,Hartford2007PretrialIllness,Cosden2003EvaluationTreatment,Lamb1996CourtOffenders}. Although evidence around the relative short-term costs and savings has been considerably mixed, depending in great degree on the implementation details and variation in costs of incarceration across communities \cite{Cowell2004TheAbuse,Steadman2005AssessingDisorders}, there seems to be a growing consensus that diversionary programs that reflect individuals' specific challenges and needs can have a positive impact on those individuals.

\subsection{Our Work}

This paper describes a collaboration 
\begin{anonsuppress}
between the University of Chicago's Center for Data Science and Public Policy\footnote{The corresponding authors and the collaboration have now moved to Carnegie Mellon University} and the Los Angeles City Attorney's Office 
\end{anonsuppress}
to develop individualized intervention recommendations (i.e., diversions, conditional plea agreements, stayed sentencing, or other favorable case disposition based on appropriate social service linkage rather than traditional sentencing methods) by identifying individuals most at risk for future arrests for misdemeanor offenses handled by their office. The case study we present here is focused on dealing with equity, fairness, and bias issues that come up when building such systems, including: identifying desirable equitable outcomes from the policy view, defining these metrics for specific problems, understanding their implications on individuals, performing machine learning model development and selection, and helping decision-makers decide how to achieve their policy outcomes in an equitable manner by implementing such a system. While there has been a lot of theoretical work done on fairness in machine learning models in resource allocation settings, our work is focused on taking the many definitions and metrics for fairness that exist in literature, and showing how to operationalize those definitions to select a metric that optimizes a specific policy goal in a public policy problem. \textit{We believe that this mapping from theory to practice is critical if we want data-driven decision making to result in fair and equitable policies.}


The ethical implications of applications of machine learning to criminal justice systems, particularly recidivism risks, has been the subject of considerable work and recent debate. The May 2016 publication by ProPublica of an investigation into the predictive equity of a widely-used recidivism risk score, Correctional Offender Management Profiling for Alternative Sanctions (COMPAS), helped raise both public awareness and researcher interest in these issues. Their analysis found dramatic racial disparities in the score's error rates, with false positive rates nearly twice as high for black defendants relative to white defendants and false negative rates roughly twice as high for white defendants, despite similar levels of precision across racial groups \cite{Angwin2016MachineBias,Larson2016HowAlgorithm}. Subsequent scholarly work further explored the COMPAS example as well as the theoretical limitations of various competing metrics for measuring fairness \cite{Hardt2016EqualityLearning,Chouldechova2017FairInstruments}. More recently, Picard and colleagues \cite{Picard2019BeyondFairness} used anonymized data from New York city to demonstrate the generalization of ProPublica's findings to another context and explore more equitable options for implementing risk assessment in bail determination.

\subsection{Machine Learning in Criminal Justice}

The ongoing debates about both the context-specific definitions of fairness and the implications of not being able to meet all definitions at the same time are far from settled, and researchers continue to explore these topics in both the machine learning and legal literature. While some (such as Picard and colleagues \cite{Picard2019BeyondFairness} as well as Skeem and Lownkamp \cite{Skeem2016RiskImpact}) see the promise of algorithms carefully designed with equity in mind to improve on a status quo rife with subjectivity and biases, others raise questions about the practical ability of these tools to overcome existing disparities in the criminal justice system. Citing concerns about biased input data and conflicting definitions, Mayson \cite{Mayson2019BiasOut} argues for restraint in the use of any predictions in criminal justice applications, particularly for punitive outcomes such as denying bail or handing down harsher sentences. Likewise, Harcourt \cite{Harcourt2015RiskAssessment} argues that strong associations between prior arrest history and race could exacerbate the ``already intolerable racial imbalance'' in prison populations through the growing use of risk scores in criminal sentencing.

While our work focuses on an \textit{assistive intervention} use case of identifying at risk individuals for social service interventions that seem to raise fewer inherent ethical concerns for many authors (e.g. Mayson \cite{Mayson2019BiasOut} and Harcourt \cite{Harcourt2015RiskAssessment}), we nevertheless believe it is important to carefully consider fairness in these predictions in order to ensure that scarce resources are being allocated in a manner consistent with social goals of fairness and equity, instead of purely optimizing for efficiency alone. Ideally, to the extent that these programs may lower the risk of future arrests associated with individuals' existing challenges, accounting for predictive fairness in programs that help divert individuals from jail may even help counterbalance existing disparities in incarceration rates of these vulnerable populations.

Previous work has enumerated metrics for evaluating bias \cite{Verma2018FairnessExplained,Gajane2018OnLearning}, explored inherent conflicts in satisfying them \cite{Hardt2016EqualityLearning,Chouldechova2017FairInstruments}, and described case studies and applications to a variety of problems \cite{Chouldechova2018ADecisions,Hardt2016EqualityLearning,Rajkomar2018EnsuringEquity.,Beutel2019PuttingImprovements}. The main contributions of this work include our framework for equity analysis, methods for balancing equity with other goals such as efficiency and effectiveness, and the application of this framework and methods to a public policy problem. Section 2 discusses the context of the work, data, and our approach. Section 3 briefly reviews the results of modeling and initial validation on novel data. Section 4 explores the potential sources of bias in this context while Section 5 discusses predictive fairness specifically and strategies for mitigating disparities. Section 6 concludes and discusses implications for similar applications and opportunities for future research.

\section{Problem and Approach}

\subsection{Recidivism Reduction in Los Angeles}
The Los Angeles City Attorney's Office has taken a leading role in developing and implementing innovative programs to improve individual outcomes and public safety. Their array of community justice initiatives reflect principles of partnering with the community to work in its best interest, creative problem solving, civic-mindedness, and attorneys embodying a leadership role in the community \cite{Feuer2019CommunityJustice}. Many of these programs have received recognition for their holistic view of justice and the City Attorney's role in the community, including pop-up legal clinics for homeless citizens, prostitution diversion efforts, and a neighborhood justice initiative that focuses on restorative justice over punitive responses for low-level offenses \cite{BeverlyPressStaffReporters2017LosOffice,Wagman2016Op-Ed:Offender,Rothstein2016NeighborhoodInvolvement}.

Believing that traditional prosecutorial approaches have proven insufficiently effective as a response to misdemeanor crime --- particularly in the context of a city facing overcrowded jails, endemic homelessness, and closures of county courthouses --- the LA City Attorney has also recently created the Recidivism Reduction and Drug Diversion Unit (R2D2) to develop, oversee, and implement new criminal justice strategies rooted in evidence-based practices, data analytics, and social science. The unit has seen success with proactive community outreach programs (such as LA DOOR \cite{BSCCStaff2019OutreachGrant}) seeking to bring services to, and remove legal barriers from, individuals afflicted with substance abuse, poverty, and homelessness. But R2D2 also has a more ongoing role as well, seeking to improve the results of individuals who frequently cycle in and out of the criminal justice system as they show up involved with new misdemeanor cases.

Recognizing that these chronic offenders reflect a failure of the existing criminal justice system to either deter future offenses through punitive actions or improve the underlying challenges that are leading the individual back into the system, R2D2 aims instead to develop individualized social service intervention plans in hopes of disrupting this unproductive cycle. However, the unit faces a number of challenges in preparing such diversion plans in real time when a case arises: the heavy caseload handled by the City Attorney's Office, very short turn-around times between initial booking and prosecutorial resolution, and poor data integration (including, in some cases, paper records). Ideally, these intervention plans could be prepared in advance and ready for implementation if and when a given individual was seen by their office again. However, because the process of developing case histories and recommendations for appropriate social service interventions is time and resource intensive, R2D2 could not practically prepare them for the large number of individuals who have been involved in past cases and instead needs a means of prioritizing the individuals most likely to be involved in a new misdemeanor case in the near future in order to effectively implement such a program.

To aid R2D2 in identifying chronic offenders, determining caseload priorities, developing prioritized interventions, and protecting public safety, the Los Angeles City Attorney partnered with us to develop predictive models for the risk of a given individual to be involved with a subsequent interaction with the criminal justice system. The goals of this work were to build a system that 1) enables efficient use of the limited resources the City Attorney's office has, and 2) results in mitigating existing disparities in criminal justice outcomes.

\subsection{Data}
Data extracts from the City Attorney's case management system were provided for the project. As with any project making use of sensitive and confidential individual-level records, data protection is of the utmost importance here and all the work described in this paper was done under strict data use agreements and in secure computing environments. These data included information about jail bookings, charges, court appearances and outcomes, and demographics relating to cases handled by their office between 1995 and 2017. Because the system lacks a global unique person-level identifier, case-level defendant data was used to link cases belonging to the same person using a probabilistic matching (record linkage) package, \verb|pgdedupe| \cite{Bauman2018ImprovingIdentities}. Matches using first and last name, date of birth, address, driver's license number (where available), and California Information and Identification (CII) number (where available) identified a total of 1,531,534 unique individuals in the data, associated with 2,456,365 distinct City Attorney cases.

\subsection{Machine Learning Modeling Strategy and Goals}
\label{sec:goals}
To assist R2D2 with their workload management and proactive case and intervention preparation, we used these data to develop predictive models of individuals likely to cycle back into the criminal justice system, choosing as our target variable (label) an indicator of whether a given individual was associated with at least one new booking into the local jail or City Attorney case in the subsequent six months. It is worth highlighting that, as several authors have noted previously \cite{Angwin2016MachineBias,Chouldechova2017FairInstruments,Mayson2019BiasOut,Kroll2016AccountableAlgorithms,Harcourt2015RiskAssessment}, target variables focused on subsequent arrest, booking, or prosecution are highly imperfect proxies for subsequent crime commission (because, particularly for lower-level offenses, not all crimes committed lead to arrests, and policing practices and decisions may result in disparities between communities in enforcement rates), nor can or should the resulting scores be interpreted as any reflection of the underlying criminality of the individuals about whom predictions are made. We suggest that two factors mitigate these potential ethical concerns in this case: First, that the nature of this program is supportive and designed to help the individual rather than punitive ameliorates the potential for harm associated with being predicted to have a high risk. And, second, the reactive nature of the intervention means that predicting the likelihood of subsequent interaction with the criminal justice system is in fact the appropriate outcome of interest here: the tailored intervention plans will only be put into effect for those individuals who are involved in a subsequent case handled by the City Attorney and the aim of the program is to provide better outcomes for these people if and when they do return. We do recognize that there are potential ethical issues here around the misuse of such a system when given to the wrong agency but have worked closely with the organizations involved to ensure that this does not happen.

From its inception, this work had two key goals: First, to improve the efficiency of R2D2's ability to serve the community through appropriate social service intervention programs by identifying individuals for whom advance preparation of individualized intervention plans was likely to be warranted. And, second, to ensure that the program resulted in equitable outcomes, consistent with the unit's goals of improving outcomes in traditionally under-served communities and working to mitigate existing disparities in criminal justice outcomes. As such, we sought to develop models that were effective at predicting future interactions with the criminal justice system, while evaluating the predictive fairness of these models and taking steps to ensure decisions based on these predictions were equitable as discussed further in Section \ref{sec:fairness} below.

An important assumption to make explicit here is that the additional consideration individuals will receive on a subsequent case as a result of being selected by the model will in fact accrue to their benefit (as well as enhance public safety in general) by helping them successfully exit the criminal justice system in the long run. We arrived at this assumption through the process of scoping and defining the project in detailed conversations with the City Attorney's Office, as well as our understanding of the scholarly literature surrounding the needs and barriers to success of many individuals involved in the criminal justice system. In particular, our belief that a better understanding of how the criminal justice system has failed to improve outcomes for these individuals in the past will allow R2D2 to develop forward-looking strategies that will do so in the future provides the foundation for how we analyze and understand the fairness implications of our predictive model in Section \ref{sec:fairness}. However, this assumption can and should be tested \textit{rigorously} and \textit{regularly}  in a fully implemented program and, if found to be faulty, a review of the equity and ethical implications of the work would be necessary.


Because the program was focused on improving outcomes for people frequently cycling through the criminal justice system, we focused our modeling efforts on those individuals who had more than one prior interaction (initial analyses also indicated that this cohort was far more likely to experience a subsequent interaction as well). Feature construction, model training, and performance evaluation was performed with the open-source machine learning toolkit, \verb|triage| \cite{Ackermann2018DeployingPolicy}. Features developed from the input data included information on the number and type of previous charges (structured to indicate the type and relative seriousness of each offense), information on origins and outcomes of prior City Attorney cases, demographics, prior jail bookings (and associated charges), and frequency and recency of prior criminal justice interactions. A grid of binary classification methods (including regularized logistic regressions, decision trees, random forests, and extra trees classifiers) and associated hyperparameters was evaluated for performance on the task of identifying the top 150 people most at risk of a new case or booking in the next six months, with the focus on the model's top 150 chosen as a potentially reasonable workload for R2D2. To ensure evaluation and model selection was done in a manner that reflected performance on novel data in a context in which policies and practices may change over time, we used a strategy of inter-temporal cross-validation \cite{Hyndman2018Forecasting:Practice} with modeling dates spaced at 6 month intervals between January 1, 2012 and January 1, 2017, each evaluated on the subsequent six month period.

\section{ML Modeling Results}

\begin{figure*}[ht]
  \centering
  \includegraphics[width=\linewidth]{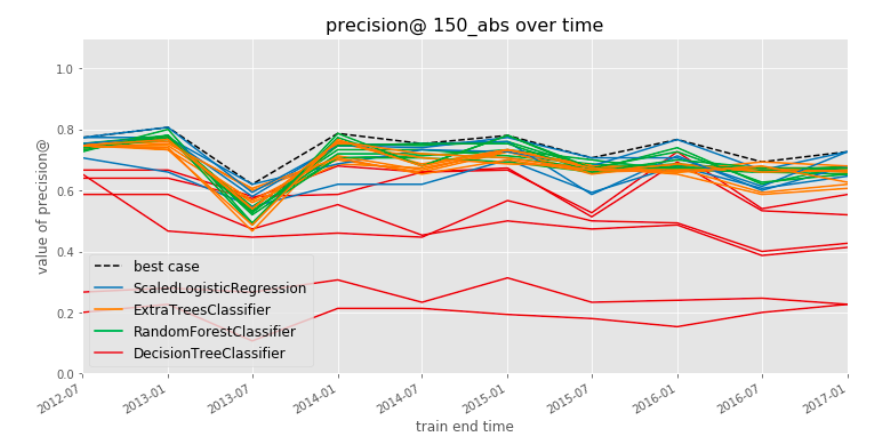}
  \caption{Inter-temporal cross-validation grid search across classification methods and hyperparameters. Each line represents the precision among the top 150 highest-risk individuals as determined by a given model specification for data up to each point in time and evaluated on the subsequent six months. Lines are colored by classification method.}
  \Description{A line graph showing results of several model specifications on the prediction of the 150 individuals most at risk of a new interaction with the criminal justice system in the subsequent six months. Many of the specifications (logistic regressions, random forests, extra trees) perform relatively similarly with precision clustering between 70-80\%, while many of the decision tree specifications perform much more poorly.}
  \label{fig:audition}
\end{figure*}

Results of the grid search used for model selection are shown in Figure \ref{fig:audition}. Many of the models and hyperparameters tested performed in a similar range, with precision (positive predictive value) at the top 150 varying over time in a range between 70-80\%, and a final model was chosen for its balance between overall performance and stability.\footnote{A random forest with 1000 estimators, a maximum depth of 50, minimum of 100 samples per split using the gini criterion, and the square root parameter for determining the maximum number of features used.}

As of January 1, 2017 the City Attorney's data included 415,614 individuals who had more than one prior misdemeanor case or jail booking (and were included in the model built at that time). The baseline rate at which these individuals had a new criminal justice interaction over the next six months was 4.4\% (18,374), indicating that relatively few people eligible to be included in the model were seen again over the evaluation period. From January 1 through June 30, 2017, 109 of the 150 highest-risk individuals identified by the model were involved with a new case or booking in this time window, a rate of 73\%, and much higher than the overall 4.4\% (random) baseline. Among the most predictive features used by this model to identify risk are the individual's age (both at time of first arrest and as of the prediction date), number of recent priors, and recency of their last interaction with the criminal justice system.

With any modeling system built on temporal data, there is always the possibility that information ``leaks'' from the future to artificially improve model performance. For example, a coding error may cause events to become \textit{misdated}. Although we diligently searched for such errors in the system, the best test of a model's performance is how well it predicts events that haven't happened yet. As a test of how the model would perform on new events that we did not have access to when we built the system, we used our modeling tools to make predictions for the second half of 2017 at the conclusion of the initial model development. In 2018, we received a second data transfer from the LA City Attorney and matched the new cases and bookings from July 1, 2017 through December 31, 2017 to our predictions for that period and found that, out of the 150 highest risk individuals, 104 (69\%) went on to have a new case or booking during the last half of 2017. Taken together, these results indicated that the predictive model we had developed could perform and generalize reasonably well at achieving the project's first goal of improving the efficiency of R2D2's efforts to proactively develop individualized diversion plans for people likely be involved in case handled by their office in the near future.


\section{Bias and Fairness}
Gathering case histories and developing individually-tailored recommendations for social service intervention plans is a time and resource intensive process for the R2D2 staff. Even if given considerable advanced warning of individuals likely to be seen by their office again in the future, they would only be able to do so for a small fraction of individuals. They therefore want to ensure that they are allocating these scarce resources in a manner that is both efficient and equitable.

As in any machine learning problem, there are a number of potential sources of bias that could influence the equitability of our results: the representativeness of the sample, accuracy of labels/outcomes and columns/variables, data reconciliation and processing, feature engineering, the modeling pipeline, and program implementation (such as the assignment and effectiveness of interventions). As other authors have discussed, particular concerns in the criminal justice context stem from sample and label biases \cite{Angwin2016MachineBias,Chouldechova2017FairInstruments,Mayson2019BiasOut,Kroll2016AccountableAlgorithms,Harcourt2015RiskAssessment}. Over-policing in communities of color may lead both to an unrepresentative sample for recidivism projects as well as label issues when subsequent arrests are used as indicators of future criminality. Likewise, racial disparities in conviction rates and sentencing may introduce bias into labels that rely on these criminal justice outcomes. A broad array of socioeconomic factors certainly contribute to historical and ongoing disparities in underlying crime rates that can inform programmatic goals and concepts of fairness even when labels may be considered reliable.

Improving machine learning results with respect to fairness has recently been a very active area of research, with several innovative approaches proposed at various stages of the process. Providing a framework for decomposing the components of biases, Chen and colleagues \cite{Chen2018WhyDiscriminatory} suggest that targeted collection of additional examples or new features may be an effective mitigation strategy in some cases. Others, including Zemel and colleagues \cite{Zemel2013LearningRepresentations}, Celis and colleagues \cite{ElisaCelis2019ClassificationGuarantees}, Edwards and Storkey \cite{Edwards2015CensoringAdversary}, Agarwal and colleagues \cite{Agarwal2018AClassification}, and Zafar and colleagues \cite{Zafar2017FairnessClassification,Zafar2017FairnessMistreatment} have focused on accounting for biases directly in the learning process by making modifications such as introducing costs for departures from equity into the loss function during model training. Equity metrics have also been introduced in the process of model selection \cite{Chouldechova2017FairerWhom,Steif2019AlgorithmicScientists}, balancing test set performance in terms of both accuracy and fairness in making the choice of modeling method and associated hyperparameters. Where an existing classifier shows disparate results, Dwork \cite{Dwork2018DecoupledLearning} described methods for eliminating biases by learning separate group-specific models on top of the existing classifier, and Hardt \cite{Hardt2016EqualityLearning} likewise describes model-agnostic post-processing steps to mitigate disparities.

Even when taking steps to account for and remove bias issues earlier in the pipeline, auditing the resulting predictions for fairness, using tools such as \verb|aequitas| \cite{Saleiro2018AequitasToolkit}, is necessary to understand both how effective these mitigation strategies have been and detect any residual biases. Our approach in Section \ref{sec:fairness} focuses on this latter phase of post-hoc bias detection and mitigation. And, although we directly use labels that reflect future interactions with the criminal justice system, we do not rely on an assumption that these labels provide an unbiased representation of subsequent \textit{criminal activity} and in fact explore an approach to predictive fairness that seeks to counteract existing base rate disparities that might arise from the sorts of sample and label biases that others have raised as potential concerns when working with criminal justice data. Additionally, in Section \ref{sec:discussion} we provide further thoughts on detecting and avoiding biases in program implementation, both in this particular case as well as more generally.

\section{Predictive Fairness}
\label{sec:fairness}

\subsection{Measuring Fairness}
Much has been written about the competing (and often mutually exclusive) concepts of fairness in machine learning problems \cite{Verma2018FairnessExplained,Gajane2018OnLearning,Chouldechova2017FairInstruments,Hardt2016EqualityLearning}. In the context of recidivism prediction, this debate has focused primarily on punitive applications, such as risk scores being used to deny defendants bail or even to assign harsher sentences to individuals with higher risk. In that setting, individuals may be harmed by being predicted to be at higher risk than they in fact are: that is, many of the relevant fairness metrics include some measure of false positives produced by the score.

The program we focus on in this work, however, is supportive in nature, aiming to improve long-term outcomes for defendants through diversion programs, tailored social service interventions, and additional consideration of their case history (refer to Section \ref{sec:goals} for a discussion of the underlying assumptions here). Moreover, because the tailored intervention recommendations will only be acted upon on a subsequent case, the interventions only apply to individuals who the model correctly classifies as high risk. As such, there is minimal risk of individual harms accruing from false positives (while they do represent wasted effort on the part of the R2D2 team, we see relatively few equity considerations in that regard). Instead, the individuals who could be viewed as harmed by an inequitable application of this program are those who might have benefited but were mistakenly classified as unlikely to return: that is, the model's false negatives.

\begin{figure*}[htbp]
  \centering
  \includegraphics[width=\linewidth]{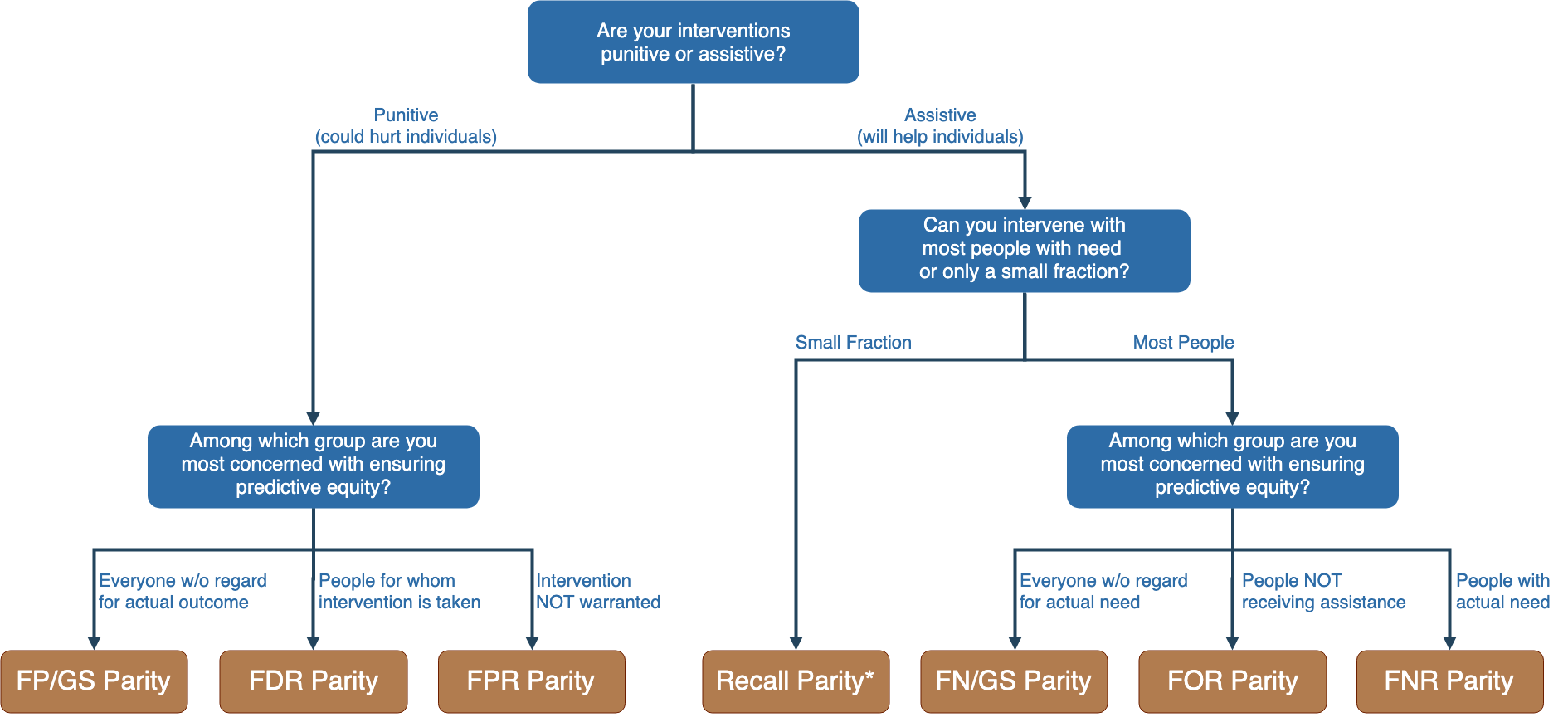}
  \caption{A framework for considering potential fairness metrics. Here, FP/GS is the number of false positives divided by the total size of each group of interest (e.g., the subset of individuals of a given race, gender, etc), FN/GS is the analog with false negatives, FDR is the false discovery rate, FPR the false positive rate, FOR the false omission rate, and FNR the false negative rate. In practice, considering the trade-offs across multiple metrics is often desirable. Note that while focusing on recall in the case of a small assistive program is equivalent to focusing on FNR parity, it may have nicer mathematical properties, such as meaningful ratios.
  }
  \Description{A tree diagramming the authors' process for selecting predictive fairness metrics based on the nature of the program and group among who ensuring predictive equity is of concern. For punitive programs, a focus on everyone without regard to actual outcomes would lead to FP/GS Parity, while a focus on people for whom intervention is taken would lead to FDR parity, and a focus on people for whom intervention is not warranted would lead to FPR parity. For assistive programs that can only serve a small fraction of need, we focus on recall parity. For larger assistive programs, a focus on everyone without regard to need leads to FN/GS parity, while a focus on people not receiving assistance leads to FOR parity, and a focus on people with actual need leads to FNR parity.}
  \label{fig:fairnesstree}
\end{figure*}

In most cases, this would lead us to consider equity metrics that focus on disparities concerned with individuals who may benefit from the assistance but are left out from the program, such as the false omission rate or false negative rate (Figure \ref{fig:fairnesstree} provides more detail on our framework for choosing predictive fairness metrics). However, the limited scale of the program due to the office's constrained resources poses additional challenges for thinking about equity. Because intervention recommendations can only be prepared for a small fraction of the individuals who will actually be charged with another misdemeanor, any implementation will unavoidably have a large number of false negatives.

A focus on false omission rate parity, for instance, in not meaningful for such a small program because the false omission rates will very nearly approximate the underlying prevalence for each group and not be possible to balance given the limited number of people who can received assistance. Likewise, in these cases, the false negative rate for each group will be very close to 1 --- although balancing $FNR$ across groups in these cases is possible, focusing equivalently on recall is easier in practice (for instance, with more meaningful ratios across groups). Additionally, in the case of limited resources, we see a reasonable interpretation of recall as fairness metric in itself, noting that it corresponds to what Hardt and colleagues \cite{Hardt2016EqualityLearning} term ``equality of opportunity'': given that the program cannot serve everyone with need, we may want to at least ensure that the set of people it does serve is representative of the distribution of need across protected classes in the population.

\begin{figure}[ht]
  \centering
  \includegraphics[width=0.8\linewidth]{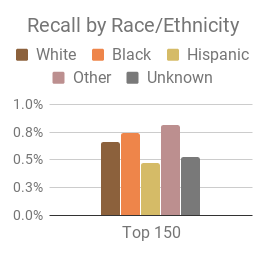}
  \caption{Recall among the 150 highest-risk individuals by race/ethnicity in the selected model, showing an under-representation of hispanic individuals (as well as individuals with unknown race/ethnicity) in this set.}
  \Description{A bar graph showing recall among the 150 highest-risk individuals from the selected model: 0.66\% for white individuals, 0.74\% for black individuals, 0.47\% for hispanic individuals, 0.81\% for individuals with other race/ethnicity, and 0.53\% for individuals with unknown race/ethnicity.}
  \label{fig:recall150}
\end{figure}

To evaluate the predictive fairness of our best-performing model, we looked at the distribution of recall (also known as sensitivity) by race/ethnicity\footnote{We use race and ethnicity as a combined field in this paper because that is how the data was collected and organized in the LA City Attorney's Office system.}. Figure \ref{fig:recall150} illustrates the presence of disparities if the model were used to select the 150 highest-risk individuals without consideration of equity. While recall is similar for black and white individuals, hispanic individuals are considerably underrepresented in the top 150 group relative to their actual prevalence.

\subsection{Mitigating Disparities}
While the predictive performance of the model satisfied the goal of efficiency (using precision or positive predictive value as the metric) defined at the outset of the project, the racial disparities found above fell short of satisfying the equally important goal of fairness. In order to remedy this shortcoming, we explored the use of slightly adjusting the score threshold used by the model to select individuals from each race/ethnicity group to better balance recall across the groups.

Some authors have argued that using separate thresholds in the interest of balancing predictive equity in itself falls short of fairness by treating individuals with similar risk profiles in different ways \cite{Corbett-Davies2017AlgorithmicFairness}. However, concepts of fairness through unawareness have been consistently demonstrated to be misguided \cite{Dwork2012FairnessAwareness,Kroll2016AccountableAlgorithms,Calders2013WhyProcedures,Taslitz2007RacialJustice,Bonilla-Silva2015TheAmerica,Fryer2007AnAction}, and any process that seeks to balance the dual goals of equity and efficiency will face an inherent trade-off between these objectives, even where it is obscured by the process involved. For instance, when a more equitable but less predictive model is chosen over a more predictive but less equitable one to distribute a benefit, there will always be some individual whose score in the more predictive model would have qualified them for a benefit that they didn't receive as the result of choosing the more equitable model. Though both models may be well-calibrated in a limited sense, data was available to better understand the risk profile of this individual that was ignored in the interest of equity, implicitly making the same trade-off as allowing the threshold to vary by group.

For further discussion of this ongoing debate from a legal perspective, see the informative pieces offered by Kroll and colleagues \cite{Kroll2016AccountableAlgorithms}\footnote{In particular, the discussion offered in Part III of their article} as well as Bent \cite{Bent2019IsLegal} and Huq \cite{Huq2019RacialJustice}, which highlight several of the competing standards and interpretations of colorblindness, equal protection, disparate treatment, and disparate impact, and their implications for algorithmic decision making. Of particular interest here is the suggestion by Kroll \cite{Kroll2016AccountableAlgorithms} that the Supreme Court's findings in \textit{Ricci v DeStefano} might prohibit any post-hoc algorithmic adjustments made in the interest of fairness along the lines of protected attributes.\footnote{Although this case was decided in the context of Title VII employment law, authors such as Kroll \cite{Kroll2016AccountableAlgorithms} and Kim \cite{Kim2016Data-DrivenWork} have looked to it to explore the more general principles the court might apply to discrimination cases more broadly.} Several others \cite{Kim2016Data-DrivenWork,Kim2017AuditingDiscrimination,MacCarthy2017StandardsAlgorithms,Bent2019IsLegal}, however, disagree with this interpretation, noting that the harm involved in \textit{Ricci} was in undoing a benefit that had already been awarded, not anything inherent in auditing or improving an algorithm after the fact, so long as those improvements are used to make future decisions rather than to reverse past ones. Other authors \cite{Sun2019TheRemedies,Lipton2018DoesDisparity} likewise speak to the potential necessity of differential treatment to avoid or mitigate disparate outcomes for ML-aided decision making. This need may even be more acute in contexts where there may be a compelling social goal of counteracting existing disparities or historical inequities. Huq \cite{Huq2019RacialJustice} further discusses the tension between between existing legal and technical concepts of fairness, suggesting a need for practical evaluation of algorithms on the basis of their actual long-term impact on disparities. Finally, from a more technical perspective, Hardt and colleagues \cite{Hardt2016EqualityLearning} make a strong case for the ability of post-processing to achieve several definitions of fairness and describe the procedure they propose as shifting the burden of uncertainty from the protected class to the decision maker. Dwork and colleagues \cite{Dwork2018DecoupledLearning} likewise explore ``decoupling'' methods that allow for improving equity by learning group-specific classifiers built on top of existing ``black box'' algorithms.


We could further consider the trade-offs involved with meeting the goal of equity in two ways:
\begin{enumerate}
    \item One option would be to measure an ``additional cost of equity'' in terms of programmatic resources. If more resources are available (or could be obtained), the scale of the program could be expanded to serve the 150 highest-risk individuals along with additional high risk individuals who are under-represented in this set.
    \item If, however, the program has a hard constraint on resources, then there is a more explicit trade-off between equity and efficiency. In this case, some individuals from over-represented groups would, of necessity, be left out in order to serve slightly lower risk individuals from under-represented groups.
\end{enumerate}

In either case, we also wanted to consider how adjusting for predictive equity might affect longer-term outcomes, particularly in the presence of underlying disparities in the baseline prevalence across groups. Assuming the program is equally effective across individuals (an assumption that does need to be validated), simply balancing recall (or sensitivity) across groups would aim to improve outcomes proportionally across groups without increasing disparities (as could happen if the model were deployed without consideration of predictive equity), but wouldn't serve to counteract existing disparities. We therefore provided an additional set of options for the City Attorney's Office to consider, balancing recall not equally across groups, but relative to their current rate of having repeated interactions with the criminal justice system. While both options will focus more resources on groups with higher need, the latter seeks to improve outcomes more rapidly for these groups relative to others, ideally resulting in equal recidivism rate across groups over time.

Because recall is monotonically increasing with the depth traversed into a score, we could readily determine thresholds that balance this metric across groups (either equally or relative to prevalence as noted above) using the procedure described in Algorithm \ref{alg:balancerecall}. For forward-looking predictions, the within-group list sizes $k_g$ were determined by balancing recall to the specific objective for the most recent complete test set.

\begin{algorithm}[ht]
    \SetAlgoLined
    \KwData{A set of test set examples with known labels $Y_i$, score $S_i$, and group membership $G_i$}
    \KwResult{The top $k_g$ number of examples to choose from each group $g$}
    For each group $g$, calculate the total number of individuals $N_g$, total number of positive labels $Y_g$ and prevalence $P_g$\;
    Sort the set of examples in each group by score, breaking ties randomly\;
    Calculate a "rolling" value of within-group recall $R_{g,i}$ and count $n_{g,i}$ up to and including each example $i$ in each sorted list\;
    \If{target equalized recall}{
        Combine these group-specific lists, sorting first by $R_{g,i}$ then by $n_{g,i}$\;
        Calculate a total count $m_i$ up to each example in this sorted, combined list\;
        \If{desired list size $K$ is specified}{
            set $k_g = max(n_{g,i}) \ni m_i \le K \forall g \in G $\;
        }
        \If{desired recall value $R$ is specified}{
            set $k_g = max(n_{g,i}) \ni R_{g,i} \le R \forall g \in G$\;
        }
        \Return{$k_g$}
    }
    \If{target recall by prevalence}{
        Choose a reference group against which to normalize prevalences, $G=g_{ref}$\;
        Calculate target ratios for each group relative to this reference: $r_g = \frac{P_g}{P_{g_{ref}}}$\;
        \If{desired reference recall value $R_{g_{ref}}$ is specified}{
            set $k_g = max(n_{g,i}) \ni R_{g,i} \le r_g \times R_{g_{ref}} \forall g \in G$\;
        }
        \If{desired list size $K$ is specified}{
            Initialize $x = min(R_{g_{ref}, i})$, a small step size $s$\;
            set $k_{all} = \sum_{g \in G} \left[ max(n_{g,i}) \ni R_{g,i} \le r_g \times x \right]$\;
            \While{$k_{all} < K$}{
                set $x = x+s$\;
                set $k_{all} = \sum_{g \in G} \left[ max(n_{g,i}) \ni R_{g,i} \le r_g \times x \right]$\;
            }
            set $k_g = max(n_{g,i}) \ni R_{g,i} \le r_g \times x \forall g \in G$\;
        }
        \Return{$k_g$}
    }
    \caption{Balancing Recall Across Groups}
    \label{alg:balancerecall}
\end{algorithm}

\begin{figure*}[htbp]
  \centering
 \begin{minipage}[b]{0.5\linewidth}
    \includegraphics[width=\linewidth]{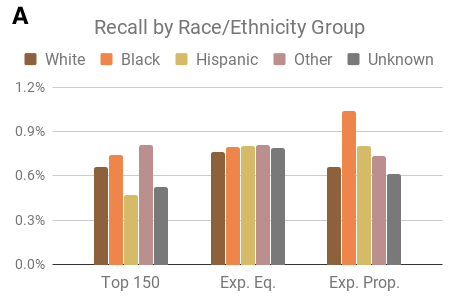}
  \end{minipage}%
  \begin{minipage}[b]{0.5\linewidth}
    \includegraphics[width=\linewidth]{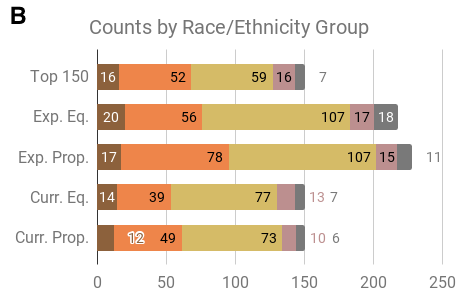}
  \end{minipage}%
  \caption{Options for balancing recall, either equally or proportional to prevalence. Options labeled "Exp." expand beyond the top 150 to include additional individuals from under-represented groups while options labeled "Curr." maintain the current scale of 150 individuals. Options labeled "Eq." equalize recall across groups while those labeled "Prop." balance recall proportional to prevalence. (A) Values of recall by race/ethnicity group in the expanded cases before and after balancing for equity. (B) Counts of individuals by race/ethnicity group for whom the R2D2 staff would prepare tailored social service intervention recommendations under each scenario.}
  \Description{Bar graphs showing the results of balancing for equity on recall and counts by race/ethnicity group. In the left panel (A), hispanic individuals are particularly under-represented when the top 150 individuals from the model are chosen without regard to equity. Balancing for equalized recall yields similar values across groups while balancing recall proportional to prevalence results in higher recall values for black and hispanic individuals. In the right panel (B), counts of individuals included by scenario are shown. In both expanded cases, nearly twice as many hispanic individuals are included relative to the top 150 (107 vs 59) and when balancing recall relative to prevalence, about 50\% more black individuals are included (78 vs 52). When limiting to the current scale, many more hispanic individuals are included (77 and 73 for equalized and proportional cases, respectively, vs 52) while fewer black individuals are included when recall is balanced equally (39 vs 52) and a similar number (49) when it is balanced to prevalence.}
  \label{fig:equitymenu}
\end{figure*}

Considering first options that expand the scale of the program in the interest of recall equity, we looked at how many additional case histories and intervention recommendations the R2D2 staff would need to be able to prepare to include the 150 highest-risk individuals as well as enough individuals from groups under-represented by this set such that either (a) every group had a recall as near to 0.81\% (the highest observed in the top 150) as possible, or (b) the ratio between the recall for each group and that for white individuals (0.66\%) was equal to the ratio of their prevalences. In the latter case, this required targeting higher values of recall for black (1.04\%) and hispanic individuals (0.80\%) relative to white individuals, as shown in Figure \ref{fig:equitymenu}A. In either case, the scale of the program would need to expand by about 50\% in order to meet these criteria: to 218 individuals for equalized recall or 228 individuals for recall balanced relative to prevalence. Figure \ref{fig:equitymenu}B breaks these counts down by race/ethnicity groups.

Alternatively, smaller thresholds can be applied to each group to satisfy these criteria on the distribution of recall while maintaining a total program size of 150 individuals. Figure \ref{fig:equitymenu}B shows how these options break down by group as well. In particular, note that many more hispanic individuals are included in either case than simply focusing on the 150 highest-risk individuals. When equalizing recall, fewer black individuals are included than even in top 150 case, while their higher underlying prevalence results in a more similar number being included when balancing recall relative to prevalence. While keeping the scale fixed at 150, we can also consider the explicit trade-offs between equity and efficiency. In this case, we find only a modest decrease in precision is required to achieve more equitable predictions: precision for both recall balanced options is only 2 percentage points lower than for the 150 highest-risk individuals without accounting for fairness (70.7\% vs 72.7\%).

Although we can explore a variety of options and make explicit the trade-offs inherent to balancing program size and costs, efficiency, and equity, the choice of how to weigh these factors against one-another is fundamentally one of policy and judgment. In practice, this involved a series of detailed conversations between the data science team 
\begin{anonsuppress}
at the University of Chicago 
\end{anonsuppress}
and the policy makers at the LA City Attorney's Office about how to understand the meaning of each metric, possible limitations of the data, available resources, and goals both for R2D2 generally and this project specifically. Not only has this process greatly helped us refine our understanding of operational predictive fairness for policy problems, but we believe it will yield better and more equitable outcomes for Los Angeles. 


\section{Discussion}
\label{sec:discussion}
As of this writing, the City Attorney's Office is implementing the system internally and exploring deploying these predictive models in their current workflow. While the model could be deployed as a ``black box'' process that periodically generates predictions based on the current state of their data, this sort of implementation runs the medium-term risk of degraded performance (in terms of both precision and fairness metrics) as patterns in the data change with changing laws and social context. Instead, an effective implementation will require ongoing evaluation of both the performance and fairness of the model's predictions over time, revisiting the model training and selection process as needed to ensure it continues to reflect changes in the underlying relationships.

From a measurement perspective, one simplifying feature of the program discussed here is its reactive nature: while social service intervention recommendations would be prepared for the set of individuals selected by the model, interventions will only take place in response to a subsequent case involving these individuals. As a result, the relevant pre-intervention outcomes for all individuals can in fact be measured, allowing for ongoing assessment of both model performance and equity. In many programs, however, this may not be the case. Where interventions are seeking to prevent the adverse outcome the model is working to predict, it may be difficult or impossible to measure true and false positives without an understanding of the counterfactual of what would have happened in the absence of the intervention.

For instance, among a cohort of unemployed individuals who receive assistance through a job training program and subsequently find employment, it would be impossible to say who would have found a job without the help of the program, inhibiting the accurate measurement of recall (along with many other potential metrics) as a means to assess performance or fairness. Data scientists and policy makers working in such contexts will need to carefully consider a strategy for ongoing measurement and feedback depending on the practical and ethical considerations relevant to their specific context, potentially drawing on methods from program evaluation and causal inference. Despite the challenges, continuing to assess and improve both efficiency and equity over time is a critical element of any predictive system that will be deployed to an ongoing application.

Finally, we should comment on the interaction between predictive fairness and fairness in outcomes. Although our focus here has been on the machine learning aspects of a project and considerations around fairness in the decision of who will receive the benefits given limited resources, this work cannot be divorced from broader questions of fairness in the context of the overall program implementation. Here, the inclusion of scenarios that incorporate disparities across racial/ethnic groups in the underlying prevalence of a subsequent interaction with the criminal justice system in the decision making process represents one step in moving beyond a simplistic view of predictive equity.

However, as programs such as the one described here are implemented, equity needs to be considered not only at the level of the machine learning pipeline, but in the context of programmatic outcomes as well. Ensuring fairness in decisions made with the aide of predictive models is an element of this broader goal of fairness in outcomes, but is far from sufficient to ensure it. In order to do so, programs need to assess the potential for differential impact of their interventions across protected groups and feed this understanding back into both their decision making about who receives interventions and, importantly, into the design of the interventions themselves to ensure they are best serving vulnerable populations.

The appropriate concept of fairness, both in decision making and implementation, is highly dependent on the nature of the program in question. The supportive nature of the social service intervention plans and implementation details of the program described here led us to focus on balancing recall in the predictive outputs of the model we developed, but this decision would be less appropriate for measuring fairness in other settings. Our hope is that the framework in Figure \ref{fig:fairnesstree} will help  machine learning practitioners and other stakeholders arrive at the appropriate concept of predictive fairness in their specific context. Likewise, as discussed above, there are ethical implications of how predictive scores such as those developed in this case study are used and interpreted. The potential for selection and label biases in the training data mean it would be highly inappropriate to interpret the resulting scores as any reflection of the underlying criminality of the individuals about whom predictions are made, let alone take any actions that reflect such an interpretation.

A related concern might involve the possibility of stigma or stereotyping associated with being identified as high risk for a future arrest. Similar issues have been described in the context of educational programs aimed at predicting students at risk of dropping out \cite{Ekowo2016TheProgram} and seem particularly salient in the criminal justice context as well. Structurally, two factors may help reduce these risks here: First, that the intervention here only involves acting on social service plans should an individual in fact be involved in another case rather than proactively reaching out to these individuals and alerting them that they have been flagged as at risk. And, second, that the intervention plans reflect what the City Attorney's Office ideally would prepare for every case (time and resources permitting) rather than a specific program developed for these high-risk individuals that might garner some stigma. Nevertheless, this concern only further highlights the fact that carefully monitoring for actual improvement in outcomes and potential unintended consequences such as these is a vital aspect of the implementation of any program intending to assist vulnerable populations.

This case study with the Los Angeles City Attorney's Office is in many ways a work in progress. We have learned a great deal from the collaboration about how to approach and understand predictive equity and the trade-offs involved in implementing a public policy program. Our hope is that these lessons and insights will prove informative to others working to balance the dual goals of equity and efficiency in the application of machine learning to other problems facing government agencies.

The methods and analyses described here are most directly applicable to other resource-constrained benefit allocation problems. Such problems, of course, are found in many public policy settings: allocating food or housing subsidies, giving additional tutoring to students, identifying long-term unemployed individuals for a job training program, or distributing healthcare workers across rural communities in a developing nation. With some modification, a similar approach certainly seems applicable to other settings (for instance, where the intervention is punitive such as with inspections for hazardous waste violations or fraud detection) so long as a single equity metric can be identified which increases or decreases monotonically with a score cut-off.

Exploring the trade-offs between equity, efficiency, and effectiveness across other contexts and applications to understand the best general approaches to balancing these goals is an ongoing research interest for us. Similarly, additional research is needed to understand how to extend this work to contexts where there is a less clearly-defined choice of fairness metric (for instance, where there are appreciable costs to disparities in false positives and false negatives) or the relevant metric is not monotonically increasing or decreasing with the prediction threshold (e.g., false discovery rate). While some recently-developed methods provide considerable flexibility for optimizing for a wide variety of fairness metrics in classification (see, for instance, \cite{ElisaCelis2019ClassificationGuarantees} for both a good example in itself and overview of other methods), a number of practical challenges remain to be addressed such as adapting these methods to the common challenge of allocating limited resources and associated non-convex ``top k'' optimization problem this implies.

\begin{acks}
This project was partially funded by the Laura and John Arnold Foundation for the Civic Analytics Network and Data Driven Justice Initiative. We would also like to thank the staff of the LA City Attorney's Office, and Dan Jeffries in particular, for their support and facilitation of this work.
\end{acks}

\bibliographystyle{ACM-Reference-Format}
\bibliography{kit_mendeley_refs}










\end{document}